\documentclass[aps,pre,groupedaddress,amsmath,runinaddress,showpacs,twocolumn]{revtex4}
\usepackage{graphicx}
\usepackage{subfigure}
\usepackage{amssymb}
\renewcommand{\vec}[1]{\mathbf{#1}}

\begin{document}

\title{Collective excitations of a spherically confined Yukawa plasma}


\author{H. K\"{a}hlert}
\author{M. Bonitz}
\affiliation{Institut f\"{u}r Theoretische Physik und Astrophysik, Christian-Albrechts Universit\"{a}t zu Kiel, 24098 Kiel, Germany}



\date{\today}

\begin{abstract}
The results of a recent fluid theory for the multipole modes of a Yukawa plasma in a spherical confinement [H. K\"{a}hlert and M. Bonitz, Phys. Rev. E \textbf{82}, 036407 (2010)] are compared with molecular dynamics simulations and the exact $N$-particle eigenmodes in the crystalline phase. Simulations confirm the existence of high order modes found in cold fluid theory. We investigate the influence of screening, coupling and friction on the mode spectra in detail. Good agreement between theory and simulation is found for weak to moderate screening and low order modes. The relations between the breathing mode in the fluid theory, simulation and the crystal eigenmode are investigated in further detail.
\end{abstract}


\maketitle

\section{Introduction}
Spherical, three dimensional dusty plasma crystals (Yukawa balls) can be created in laboratory experiments by trapping dust particles inside a glass box above the lower electrode of an rf discharge~\cite{Arp2004,arppop}. They can be regarded as the analog to confined Coulomb clusters in ion traps~\cite{gilbert1988,drewsen2010} with the difference being the screened particle interaction~\cite{bonitz2006}. Recent experimental and theoretical work has been concerned with their ground state configurations~\cite{Patrick2005,cioslowski2008,Baum08,cioslowski2010}, the occurrence of metastable states~\cite{block2008,kaeding, kaehlert2008}, time dependent shell formation~\cite{kaehlert2010} and the investigation of their normal modes~\cite{ivanov3d,henning08,henning09}, for an overview see~\cite{bonitz2010rev}.

Collective modes contain information about the particle interactions and the external confinement conditions. They are of similar interest for many-particle systems as are the atomic spectra in atomic and molecular physics. By comparing with theoretical predictions experimental normal mode measurements can be used to infer, for example, the confinement parameters and the particle charge from experimental data without changing the plasma conditions~\cite{ivanov3d,melzer2d2001}. Depending on the quality of the theory this makes them a valuable non-invasive tool for experimentalists.

In this paper we study the collective excitations of a spherically confined Yukawa plasma applying three independent methods: molecular dynamics (MD) simulations, normal mode analysis and a recent continuum approach~\cite{kaehlertpre2010} based on cold fluid theory. In Ref.~\cite{kaehlertpre2010} it was predicted that a spherically trapped plasma with Yukawa interaction possesses a much richer collective excitation spectrum than its Coulomb interacting counterpart. In particular, it was found that there exists an additional class of modes characterized by a mode index $n=0,1,2,\dots$ corresponding to the number of radial nodes in the perturbed electrostatic potential inside the plasma.

The main goal of this paper is to verify these theoretical predictions. The success of the fluid theory for Coulomb systems has been demonstrated both theoretically~\cite{dubin_md96} and experimentally~\cite{drewsen2010,bollinger1993}. However, recent work on the ground state density profile of a Yukawa plasma~\cite{Christian2006, Christian2007}, on which the fluid theory is based, indicates that correlation effects tend to become more important in systems where the particle interaction is screened and the interaction becomes short-ranged. Thus, a systematic investigation and detailed comparison with exact simulations is required in order to explore the applicability limits of the fluid theory also for Yukawa plasmas. 

By comparison with first principle MD simulations we prove that the additional modes are physical reality. Based on the very good agreement between MD and experiment~\cite{bonitz2006,kaehlert2008}, we expect that they should be observable in Yukawa ball experiments as well. Performing detailed MD simulations covering a broad range of coupling and screening parameters we are able to determine the applicability range of the fluid theory: It predicts the frequencies of the low order modes with $n=0$ with good accuracy and those of the $n=1$ modes for weak to moderate screening. For $n\ge 2$ or strong screening the deviations quickly become larger. The main source of the observed deviations are correlation effects which are missing in the cold fluid theory.

Among the eigenmodes especially the breathing mode has attracted considerable attention in recent years~\cite{sheridan2006,henning08,olivetti2009}. We therefore also compare the fluid breathing mode (the lowest monopole mode) with its counterpart among the eigenmodes of the discrete $N$-particle system in the crystalline state and the results of MD simulations.

This paper is organized as follows. In Sec.~\ref{sec:review} we briefly recall the main results of the fluid theory~\cite{kaehlertpre2010} which are relevant for the present work. The simulation method is explained in some detail in Sec.~\ref{sec:MD} and the results are compared with the theory in Sec.~\ref{sec:MDcomparison}. Section~\ref{sec:crystalmodes} deals with the exact eigenmodes of the crystallized $N$-particle system and examines the relations of the breathing mode found in cold fluid theory with the crystal eigenmodes and the MD simulations. We conclude with a summary of our results in Sec.~\ref{sec:conclusion}.

\section{Review of fluid modes}\label{sec:review}
In this section we briefly review recent results for the ground state and normal modes of a Yukawa plasma in an external confinement $V(r)=m\omega_0^2 r^2/2$ to make this paper as self-contained as possible.

For Coulomb interaction the ground state density profile in the cold fluid approximation is a sphere of constant density while for Yukawa interaction the density exhibits a parabolic decay towards the plasma boundary $R(N,\kappa)$~\cite{Christian2006}. Here $\kappa$ denotes the inverse screening length. The radius must be determined from the following equation for $\xi=\kappa R$,
\begin{equation}\label{eqn:radius}
 \xi^6+6\,\xi^5+15 \left[\xi^4+\xi^3 -k_C^3(\xi+1)\right]=0,
\end{equation}
where $k_C=\kappa R_C$ and $R_C=a N^{1/3}$ denotes the plasma radius for a Coulomb system. The relevant length scale is $a=(q^2/m\omega_0^2)^{1/3}$, corresponding to the Wigner-Seitz radius in the Coulomb limit. Accurate analytical approximations for $\xi$ are available~\cite{kaehlertpre2010}.

The normal modes have been derived in Ref.~\cite{kaehlertpre2010} by linearizing the cold fluid equations. As a result the first order potential perturbation is obtained as 
\begin{equation}\label{eqn:fluidmodes}
\hat\phi_1^\text{in/out}(r,\omega)= f^\text{in/out}(r,\omega)Y_\ell^m(\theta,\varphi),
 \end{equation}
where the indices in/out denote the regions inside ($r\le R$) and outside $(r>R)$ the plasma.

The radial function inside the plasma is given by
\begin{equation}
f^\text{in}(r,\omega)\sim r^\ell\,_2F_1\left(\frac{\alpha_\ell-\delta_\ell}{2},\frac{\alpha_\ell+\delta_\ell}{2};\alpha_\ell;\frac{\kappa^2 r^2}{x_s^2}\right),
 \end{equation}
where the parameters of the hypergeometric function $_2F_1$ are
\[
\alpha_\ell=\ell+\frac{3}{2},\:\:  \delta_\ell=\sqrt{\ell(\ell+1)+\frac{9}{4}+2 {\Omega}^2},
\]
and $x_s^2=2(\Omega_p^2(0)-\Omega^2)$. We use the notation $\Omega=\omega/\omega_0$ for the scaled eigenfrequency. Further, $x=\kappa r$ and
\[
\Omega_p^2(x)=\frac{\omega_p^2(r)}{\omega_0^2}=3+\frac{\xi^2}{2}\frac{3+\xi}{1+\xi}-\frac{x^2}{2}
 \]
denotes the local plasma frequency, which is determined by the ground state density profile.

Outside the plasma the radial function takes the form of a modified spherical Bessel function,
\begin{equation}
f^\text{out}(r)\sim k_\ell(\kappa r).
 \end{equation}

The eigenfrequencies are determined by matching the two solutions at the plasma boundary. They satisfy $\Omega^2<\Omega_p^2(\xi)$, i.e. the plasma frequency at the boundary is the maximum frequency for regular normal mode solutions with a discrete spectrum. For $\Omega_p^2(0)>\Omega^2>\Omega_p^2(\xi)$ the spectrum should become continuous, for details cf. Ref.~\cite{barston1964}.

Let us now summarize the properties of the normal mode spectrum which follows from cold fluid theory~\cite{kaehlertpre2010}:

\begin{enumerate}
 \item Normal modes are characterized by three mode numbers: the radial mode number $n=0,1,2,\dots$ and the two angular mode numbers $\ell$ and $m$ with $|m|\le \ell$. Their frequencies solely depend on $n$ and the dimensionless plasma parameter $\xi$, see Eq.~(\ref{eqn:radius}).
\item The $n=0$ modes are directly related to the surface modes of the associated Coulomb system~\cite{dubin_prl91,dubin_md96, dubin_corr96}, where $\Omega_\ell^2=3\ell/(2\ell+1)$, while the remaining modes with $n\ge 1$ originate from its degenerate bulk modes with $\Omega^2=3$. Eigenmodes with $n=0$ and arbitrary $\ell$ and $n=1$ with $\ell=0,1,2$ exist for all $\xi$ whereas all other modes exist only for $\xi>\xi_{n\ell}^\text{crit}$, where
\begin{align}\label{eqn:xicrit}
 \xi_{n \ell}^\text{crit}&=\frac{1}{2}\left[ \zeta_{n \ell}+\sqrt{\zeta_{n \ell}(\zeta_{n \ell}+4)}\right],\nonumber\\
\zeta_{n \ell}&= (2n-1)(n+\ell)-4.
\end{align}
At these points the plasma frequency assumes the values
\begin{align}
 \Omega_p^2(\xi_{n \ell}^\text{crit})=(2n+1)(n-1)+(2n-1)\ell.
\end{align}
\item 
In the limit $\xi\to\infty$, corresponding to strong screening or the macroscopic limit with $N\to\infty$, the mode frequencies are known analytically and read
\begin{equation}\label{eqn:freqlimit}
\Omega_{n\ell,\infty}^2 =2n^2+(2\ell+3)n+\ell.
\end{equation}
For finite $\xi$ they must be determined numerically from Eq.~(29) in Ref.~\cite{kaehlertpre2010}.
\end{enumerate}

\section{Molecular dynamics simulations}\label{sec:MD}
\subsection{Simulation Method}
A molecular dynamics (MD) simulation allows for an investigation of the normal modes without approximations with respect to the particle interaction or their temperature and will serve as a benchmark for the fluid theory.

In order to simulate the same conditions as in the fluid approach, we consider $N$ particles interacting through a Yukawa pair potential $\phi(r)=q^2e^{-\kappa r}/r$ in an external confinement $V(r)=m\omega_0^2 r^2/2$. The corresponding Hamiltonian is given by
\begin{equation}\label{eqn:Hamiltonian}
  H=\sum_{i=1}^{N}\frac{\vec{p}_ i ^2}{2m}+\underbrace{\sum_{i=1}^{N} V(|\vec{r}_i|) + \frac{1}{2}\sum_{i\ne j}^N \phi(|\vec{r}_{ij}|)}_{U(\vec r_1,\ldots,\vec r_N)}.
\end{equation}
In dusty plasma experiments collisions with neutral particles can be an important factor for the dynamics. We, therefore, include an additional damping term and a fluctuating force (Langevin dynamics) in the equation of motion, which is given by
\begin{equation}\label{eqn:LangevinEQ}
m\ddot{\vec r}_{i}=-\nabla_i U(\vec r_1,\ldots,\vec r_N)-m\nu \dot{\vec r}_i+\vec{f}_i(t).
\end{equation}
In thermodynamic equilibrium, the friction coefficient $\nu$ and the Gaussian noise $\vec f _i(t)$ are related by $\left< \vec{f}_i^{\alpha}(t) \vec{f}_j^{\beta}(t') \right>=2 m \nu k_\text{B} T \delta_{ij} \delta_{\alpha\beta} \delta (t-t')$, where $\alpha,\beta\in\{x,y,z\}$ and $i,j\in\{1,\ldots,N\}$. The system is fully characterized by the particle number $N$, the (normalized) screening parameter $\kappa a$, and the coupling parameter $\Gamma=q^2/(a k_B T)$.

Note that the so-defined $\Gamma$ is not the same as in a macroscopic, homogeneous system because of the inhomogeneous density in the trap. Only in the Coulomb system the mean density, and thus the Wigner Seitz radius, is constant. For Yukawa systems $\Gamma$ can be regarded as a dimensionless inverse temperature rather than a coupling parameter. Here the effective coupling varies, depending on the distance from the center of the trap and the screening parameter. Effective coupling parameters which take screening into account have been analyzed for two-~\cite{hartmann2005,ott2010} and three-dimensional~\cite{vaulina2000} Yukawa systems.

We perform finite temperature simulations where the normal modes are thermally excited and no external excitation or driving mechanism is required. When the Langevin method is used ($\nu\ne 0$), equilibration of the system is ensured by friction and the stochastic forces. 

For $\nu=0$ we rescale the individual particle velocities in the equilibration phase to enforce the desired coupling parameter via the relation $\langle E_\text{kin}\rangle=3 N k_B T/2$.  Additionally, we remove any residual rotation and center of mass motion of the whole plasma. After this initial equilibration period the system is evolved microcanonically.

\subsection{Mode detection}
In order to make contact with the fluid modes we need to find a way of measuring the fluid oscillations in the MD simulation. Since the fluid modes~(\ref{eqn:fluidmodes}) are perturbations of the induced potential, we have to derive an analogous expression which is appropriate for the MD simulation. The same method was used in~\cite{dubin_md96} for a spheroidal Coulomb system.

For a general charge distribution $q n(\vec r,t)$ the solution of the (screened) Poisson equation~\cite{Christian2006}
\begin{equation}\label{eqn:screenedpoisson}
 (\Delta-\kappa^2)\phi_\text{tot}=-4\pi q n,
\end{equation}
with the boundary condition $\phi_\text{tot}(\vec r,t)\to 0$ at infinity is
\begin{align}\label{eqn:totalpot}
\phi_\text{tot}(\vec r,t)= q\int n(\vec r',t)\frac{e^{-\kappa |\vec r-\vec r'|}}{|\vec r-\vec r'|} d\vec r'.
\end{align}
We can now expand the Yukawa potential, which is the Green's function of Eq.~(\ref{eqn:screenedpoisson}) for this boundary condition, in a series of spherical harmonics according to~\cite{arfken}
\begin{align*}
\frac{e^{-\kappa |\vec r-\vec r'|}}{|\vec r-\vec r'|}=4\pi \kappa \sum_{\ell m}   i_\ell(\kappa  r_<) k_\ell(\kappa r_>) \, Y_{\ell m}^* (\theta',\phi') Y_{\ell m} (\theta,\phi).
\end{align*}
The expansion involves the modified spherical Bessel functions $i_\ell(x)$ and $k_\ell(x)$ with the familiar arguments $r_<=\text{min}(|\vec r|,|\vec r'|)$ and $r_>=\text{max}(|\vec r|,|\vec r'|)$. 

With this expression at hand we can define multipole moments of the density. This is easily achieved by using the Green's function expansion in Eq.~(\ref{eqn:totalpot}) with the result
\begin{align}\label{eqn:multimoments}
q_{\ell m}(t)= q\sqrt{\frac{4\pi}{2\ell +1}} \int n(\vec r',t)\, \hat i_\ell(\kappa, r')\,  Y_{\ell m}^* (\theta',\phi')   d\vec r'.
\end{align}
Here we have introduced
\begin{align*}
\hat i_\ell(\kappa,r')=\frac{(2\ell +1)!!}{\kappa^{\ell}} \, i_\ell(\kappa r'),\:\: \hat k_\ell(\kappa,r)= \frac{\kappa^{\ell+1}}{ (2\ell-1)!!}\, k_\ell(\kappa r),
\end{align*}
where the prefactors are chosen such that the multipole moments reduce to their usual form in the limit $\kappa\to 0$.

These definitions allow us to write the general solution for the potential, Eq.~(\ref{eqn:totalpot}), outside the region where the density is non-zero (we used $r_<$ as integration variable) as
\begin{align}
 \phi_\text{tot}(\vec r,t)=\sum_{\ell m} \sqrt{\frac{4\pi}{2\ell +1}} \,q_{\ell m}(t)\, \hat k_\ell(\kappa, r)\, Y_{\ell m} (\theta,\phi).
\end{align}

The full time-dependence of $\phi_\text{tot}(\vec r,t)$ is now contained in the multipole moments~(\ref{eqn:multimoments}). They are easily accessible in the MD simulation because the integrals in Eq.~(\ref{eqn:multimoments}) reduce to a sum over all particles. They contain the required information about the oscillations of the potential with the mode numbers $\ell$ and $m$.

The mode frequencies are extracted from $q_{\ell m}(t)$ by computing the associated power spectral density
\begin{equation}
 \mathcal{Q}_{\ell m}(\omega)=\lim_{T\to\infty} \frac{|\mathcal{F}(q_{\ell m}(t)/N,T,\omega)|^2}{T},
\end{equation}
where $\mathcal{F}(q_{\ell m}(t)/N,T,\omega)$ denotes the Fourier transform of $q_{\ell m}(t)/N$ over a finite time interval $T$~\footnote{We use a Hann window function and moderately smooth the spectra.}. Maxima of $\mathcal{Q}_{\ell m}(\omega)$ indicate frequencies at which the system supports collective excitations.

\section{MD simulation results}\label{sec:MDcomparison}
When comparing the MD simulation and the fluid theory we immediately recognize several differences. The MD simulation is performed at finite temperature while the fluid modes correspond to a system at $T=0$, i.e. infinite coupling strength. Furthermore, the fluid approach neglects correlation effects which are fully included in the simulation. These limitations must be kept in mind when comparisons between the two are made.

The spectra obtained from the MD simulations confirm the results of cold fluid theory, e.g. the existence of additional modes, but the quantitative agreement depends on several parameters and the particular mode. In the following we will, therefore, separately discuss the influence of the particle number, screening, coupling and friction on the mode spectra and compare the simulations with the fluid approach.

\subsubsection{Screening dependence}
\begin{figure}
\includegraphics[width=0.48\textwidth]{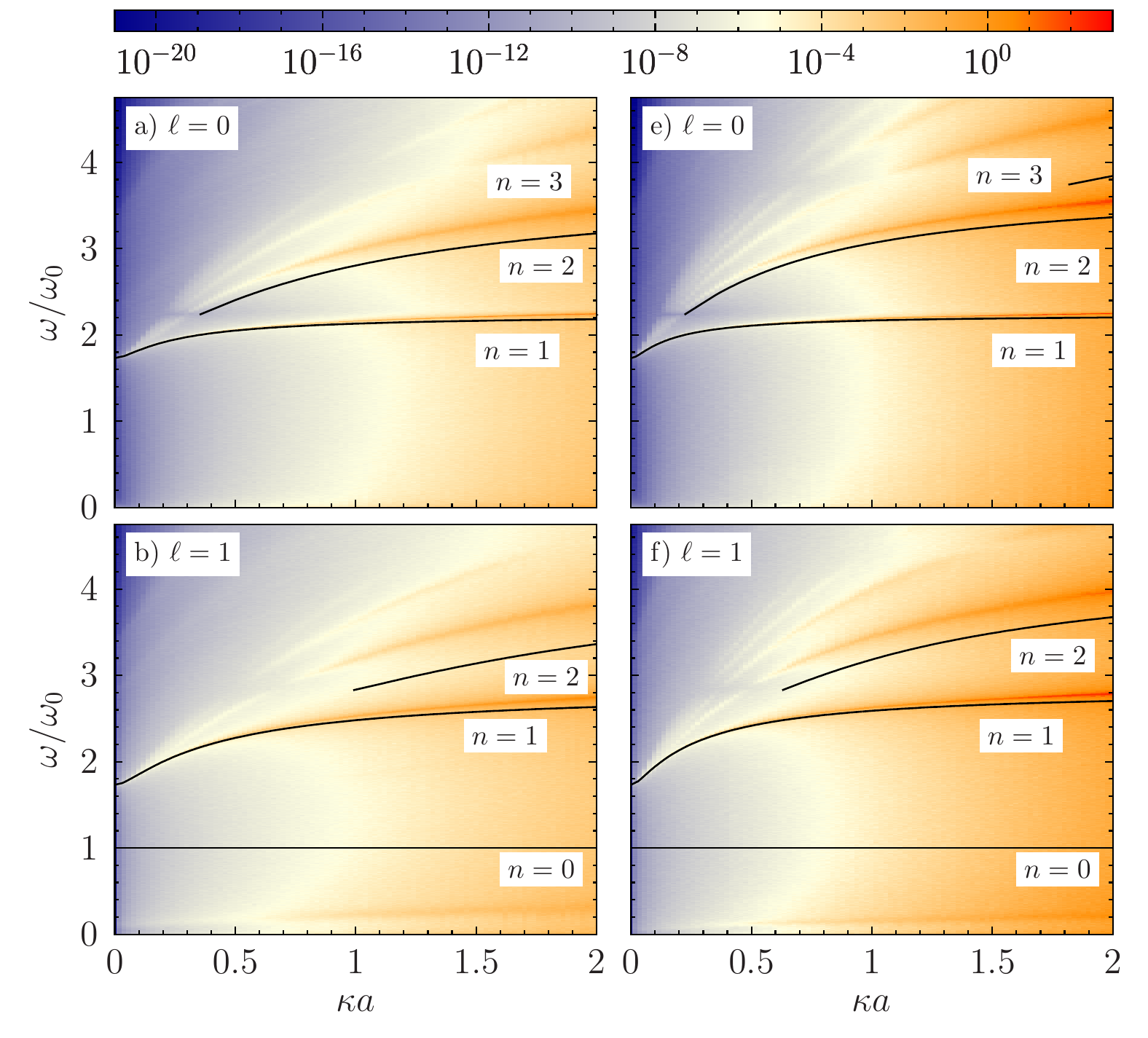}\\
\includegraphics[width=0.48\textwidth]{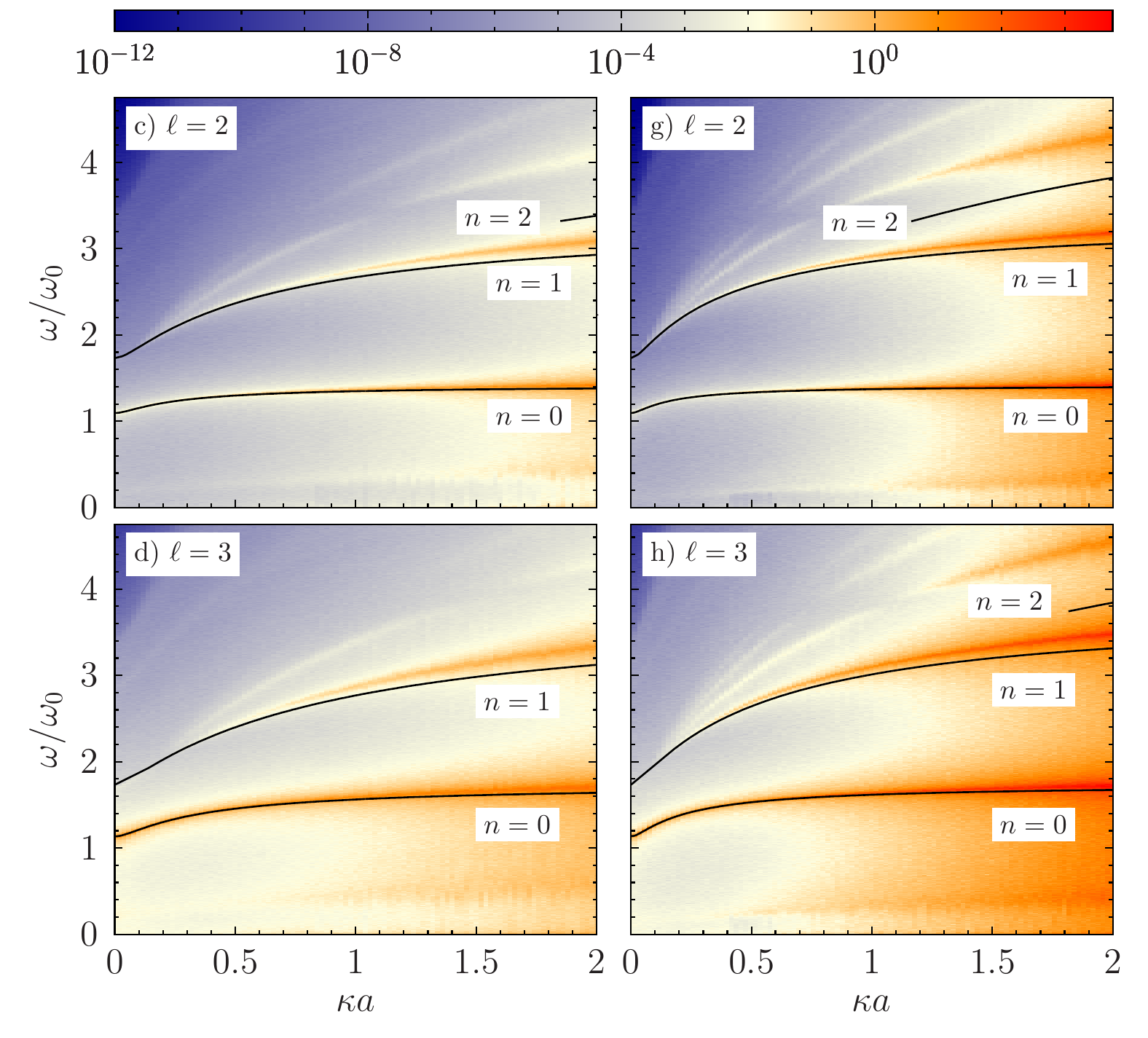}
\caption{Screening dependence of the $m=0$ multipole spectra $\mathcal Q_{\ell 0}(\omega)$ (arb. units) for various $\ell$ with $N=1000$ (left column) and $N=4000$ particles (right column) at $\Gamma=150$ without friction. The solid lines show the results of cold fluid theory.}\label{fig:N1000G150}
\end{figure}
\begin{figure}
\includegraphics[width=0.4\textwidth]{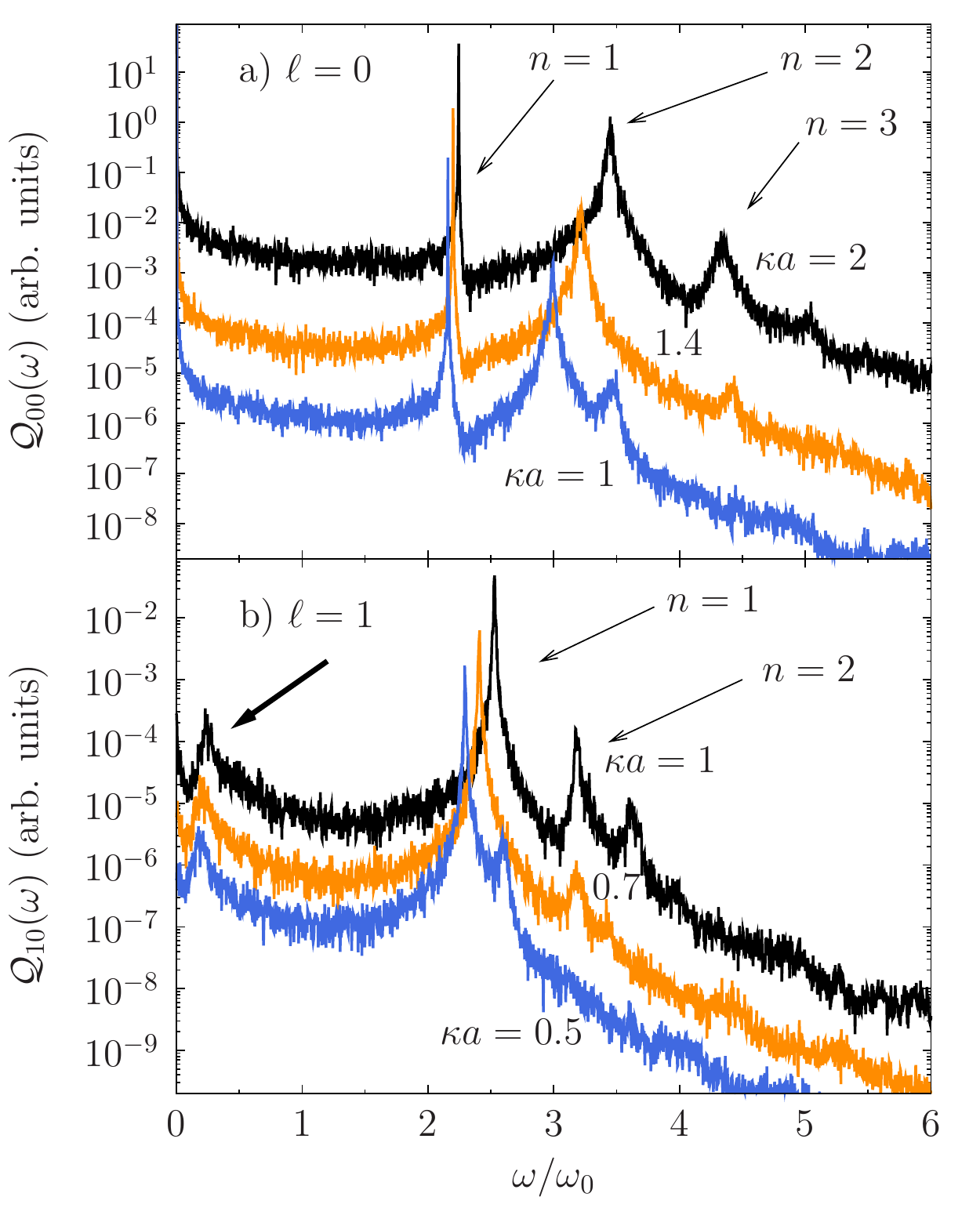}
\caption{a) Monopole and b) dipole spectra for various screening parameters with $N=1000$ particles at $\Gamma=150$. The thick arrow in b) indicates the low frequency peak in the dipole spectrum.}\label{fig:kappavar150}
\end{figure}
We begin our discussion with the influence of the screening parameter on the normal modes. Figure~\ref{fig:N1000G150} shows mode spectra for $N=1000$ and $4000$ particles at strong coupling conditions and their dependence on $\kappa a$. Note that for the same $\Gamma$ the effective coupling of the system can be quite different, depending on $\kappa a$, e.g.~\cite{ottprl2009,ott2010}. On the one hand, the interaction strength decreases as the screening parameter is increased, leading to a weaker effective coupling. On the other hand, as the interaction range $1/\kappa$ is reduced, the density increases at the same time. The cluster is being compressed which, in turn, increases the effective coupling. Thus, these two effects act in opposite directions and may at least partially cancel.

In the Coulomb limit, $\kappa a\ll 1$, we observe maxima of $\mathcal{Q}_{\ell m}(\omega)$ close to the surface mode frequencies $\Omega_\ell=\sqrt{3\ell/(2\ell+1)}$. The center of mass motion was removed in the equilibration phase so there is no maximum at $\Omega_1=1$ for $\ell=1$, see Fig.~\ref{fig:N1000G150}b. When friction is included we observe a strong peak at the expected position. Additionally, there are maxima near the plasma frequency $\Omega=\sqrt{3}$, except for $\ell=0$, where the multipole moment is constant (for $\kappa =0$ the multipole moment $\mathcal{Q}_{00}$ is simply the total charge of the cluster). The Coulomb modes have been investigated in detail in~\cite{dubin_md96}. For Yukawa interaction the general trend of our simulation results is that all mode frequencies increase with $\kappa a$ and new modes appear at strong screening which is the expected behavior from cold fluid theory. Let us now compare the results in more detail.

\textbf{Monopole modes.} For $\ell=0$ [see Figs.~\ref{fig:N1000G150}a,e] there is no mode with $n=0$. Comparing the results of cold fluid theory with the simulations we find excellent agreement for the lowest mode with $n=1$. There are only small deviations which increase with $\kappa a$. The second mode with $n=2$ exists only for $\xi>\xi_{20}^{\text{crit}}\approx 2.73$ in cold fluid theory, corresponding to $\kappa a \approx 0.35$ for $N=1000$ and $\kappa a \approx 0.22$ for $N=4000$. This result is confirmed by the MD simulation. However, there is a small upshift of the MD frequencies compared to the fluid theory which is greater for $N=1000$ than for $N=4000$. For both particle numbers we observe a third mode in the simulations at even higher frequencies which should correspond to the $n=3$ mode of cold fluid theory. However, for $N=1000$ this mode does not yet exist in the theory at the simulated screening parameters and only appears for $\kappa a>2.88$.

Since the fluid theory is a continuum approach it is expected to yield accurate results only for large $N$ and the observed discrepancies are easily understood. In particular, cold fluid theory should yield reliable results only for modes that oscillate on large length scales, i.e. for long wavelength modes. In the present case with $\ell=m=0$ the length scale of the oscillations is mainly determined by the number of radial nodes, i.e. the mode number $n$. While for $n=1$ the agreement between simulation and theory is very good, the deviations rapidly increase as we go to $n\ge 2$, due to the short wavelength character of these modes. This is qualitatively similar to the situation in a macroscopic Yukawa plasma where the Vlasov (mean field) model for the longitudinal dispersion relation $\omega(k)$ in the strongly coupled phase is only adequate for $k a\ll 1$, where $k$ is the wave number~\cite{donko2008}. Here the inclusion of correlation effects via the quasi-localized charge approximation was shown to be essential for a reliable description of the dispersion relation for larger wave numbers.

\textbf{Dipole modes.} For $\ell=1$ the center of mass mode is an exact eigenmode with $\Omega=1$, independent of $\kappa a$. Since this result is confirmed by the theory this special mode is only of minor interest here. More interesting are the higher order modes with $n=1$ and $2$ [see Figs.~\ref{fig:N1000G150}b,f]. Similar to the results for $\ell=0$ we observe good agreement for $n=1$. Again, deviations between the two approaches increase with screening. The general behavior for $n=2$ is the same. The mode exists in the MD simulation only at strong screening and the starting point is well captured by the theory. Deviations between the mode frequencies are larger than for $\ell=0$ and, again, smaller for $N=4000$ than for $N=1000$. A striking difference compared with $\ell=0$ is the existence of another low frequency mode which has no analog in cold fluid theory. Its frequency is almost independent of $\kappa a$. For $N=4000$ the frequency is lower than for $1000$ particles.

\textbf{Quadrupole and octupole modes.} In the case of $\ell=2$ and $\ell=3$ we find the first non-trivial modes with $n=0$. Here the agreement between simulation and theory is excellent. As in the previous cases the (small) deviations increase with $\kappa a$. This also holds for $n=1$. For $n=2$ the deviations between the theoretical mode frequencies and the simulation become rather large. The starting point for the existence of the modes is still well reproduced by fluid theory, at least for $N=4000$. Again, there are low frequency excitations below the $n=0$ modes which do not exist in cold fluid theory.

Inspecting the mode spectra more closely we observe another important property of the modes that exist only beyond a certain screening parameter in cold fluid theory. Left to their critical points these modes reappear in the simulations at lower $\kappa a$ for a finite screening interval, but with a relatively low intensity. This is shown in Fig.~\ref{fig:kappavar150} in more detail. In the monopole spectrum [Fig.~\ref{fig:kappavar150}a] there is a third maximum at $\Omega\approx 4.34$ for $\kappa a=2$ which is absent for $\kappa a=1.4$ but reappears at $\Omega\approx 2.99$ for $\kappa a=1$. Fig.~\ref{fig:kappavar150}b shows the same behavior for the dipole modes. Here the second peak at $\Omega\approx 3.18$ for $\kappa a=1$ is absent in the $\kappa a=0.7$ spectrum. For $\kappa a=0.5$ it reappears at $\Omega\approx 2.6$ as a second maximum rather close to the main peak. Since cold fluid theory does not allow for additional normal mode solutions below $\Omega_p(\xi)$ these excitations may be a due to the continuous part of the spectrum.

Another possible explanation could be the density profile in the MD simulation which differs quite substantially from the cold fluid result. While at the large coupling parameters we used in the simulation the exact density has a shell structure, cold fluid theory only describes the mean density correctly. Improvements in this regard are possible by including correlations via the hypernetted chain approximation~\cite{wrighton2009,wrighton2010}, see Sec.~\ref{sec:conclusion}.

\subsubsection{Influence of coupling strength}
\begin{figure}
\includegraphics[width=0.48\textwidth]{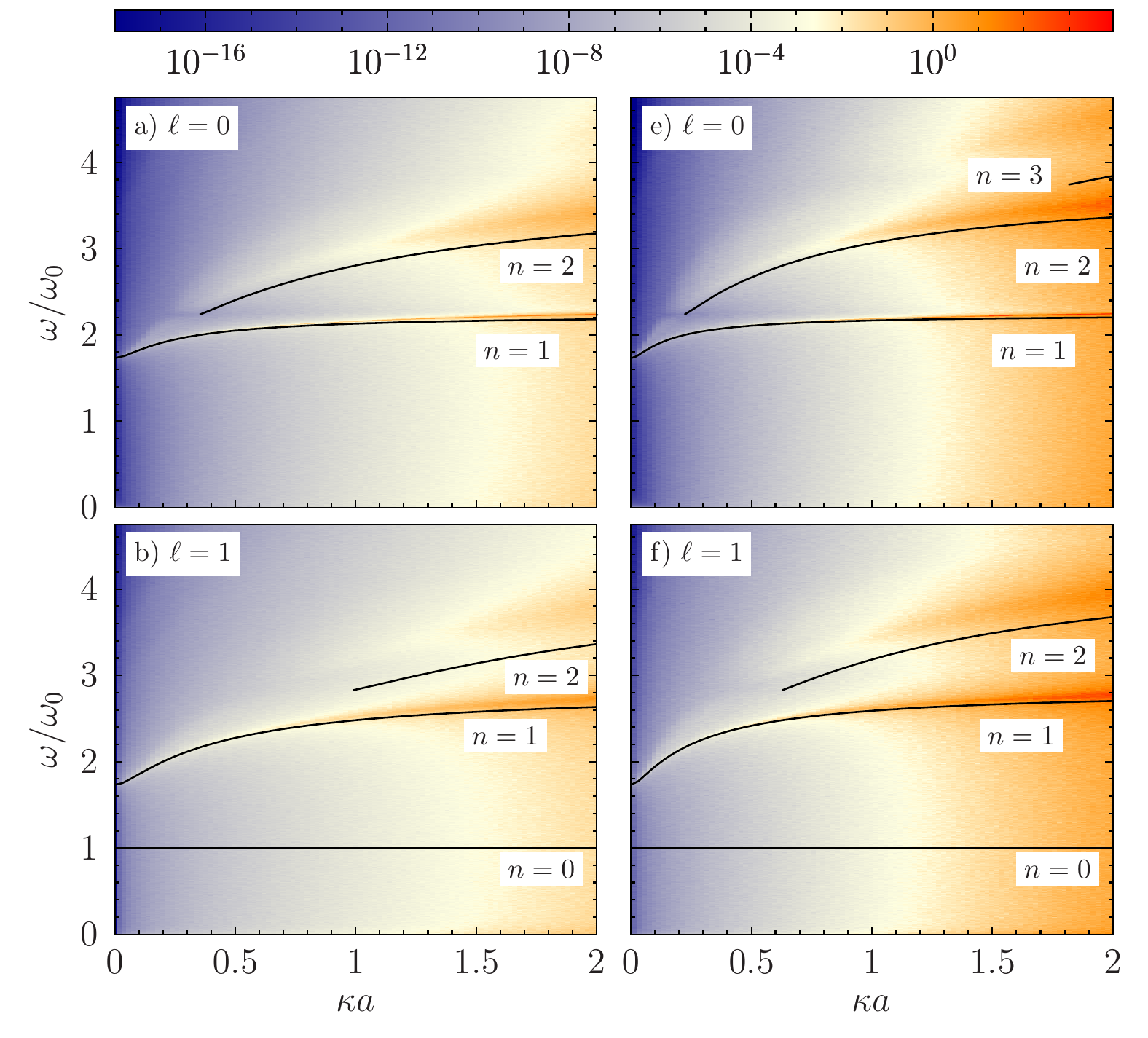}\\
\includegraphics[width=0.48\textwidth]{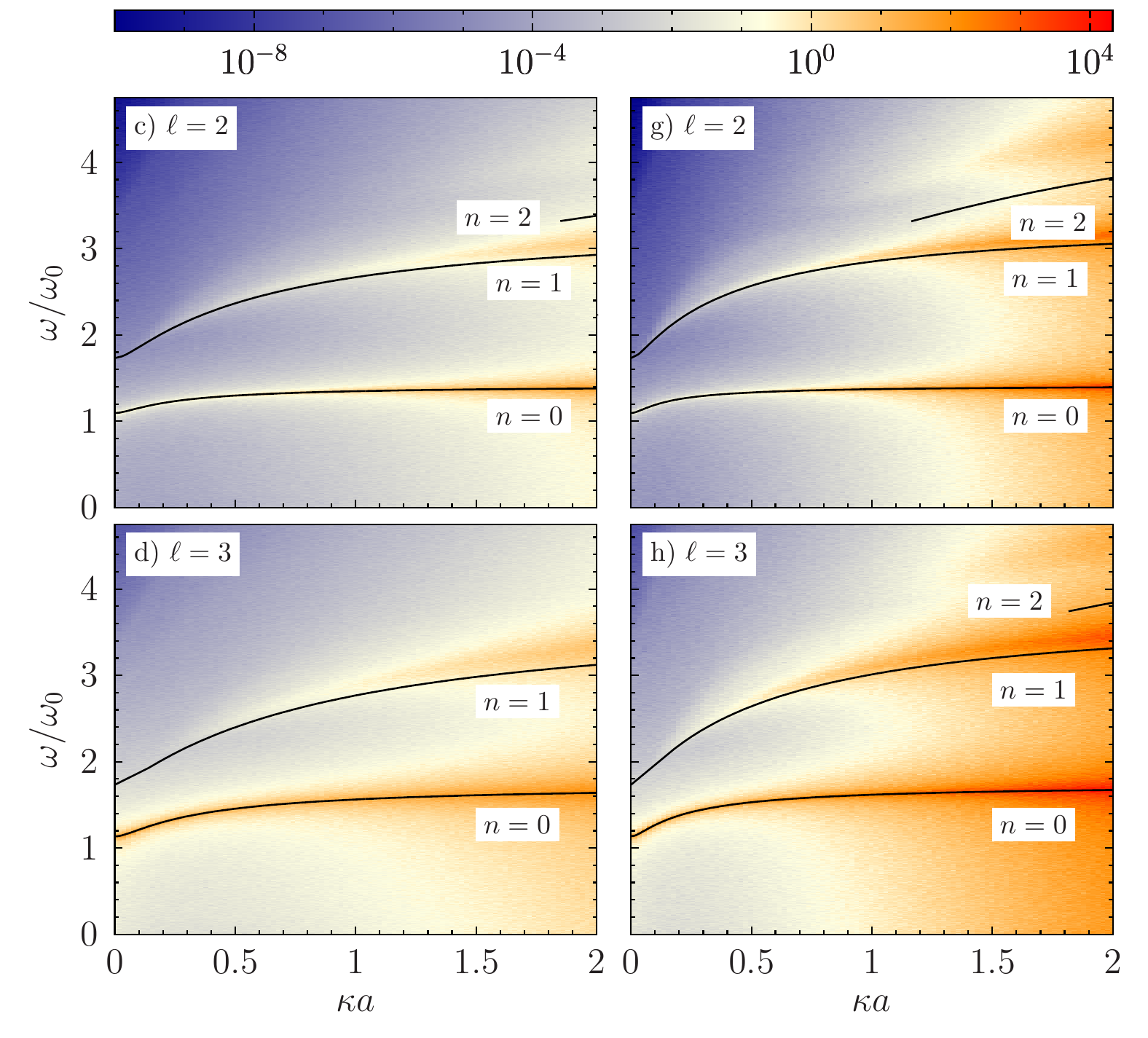}
\caption{Screening dependence of the $m=0$ multipole spectra $\mathcal Q_{\ell 0}(\omega)$ (arb. units) for various $\ell$ with $N=1000$ (left column) and $N=4000$ particles (right column) at $\Gamma=30$ without friction. The solid lines show the results of cold fluid theory.}\label{fig:N1000G30}
\end{figure}
Next, we consider temperature effects on the mode spectra. Fig.~\ref{fig:N1000G30} shows the same spectra as Fig.~\ref{fig:N1000G150} but for $\Gamma=30$. Even though the coupling is five times weaker in this case the particles are still strongly coupled. The main difference is a structural change of the plasma. For $\Gamma=150$ the density profile shows a clear shell structure while for $\Gamma=30$ the density profile is well described by the cold fluid mean field result, except for the boundary region where it smoothly goes to zero~\cite{Christian2006}. Additionally, the density has small maxima near $r=R$ which are first indications for the onset of the shell structure~\cite{dubin_corr96,kaehlert2010}. This is a correlation effect not included in cold fluid theory~\cite{wrighton2010,wrighton2009}.

\begin{figure}
\includegraphics[width=0.4\textwidth]{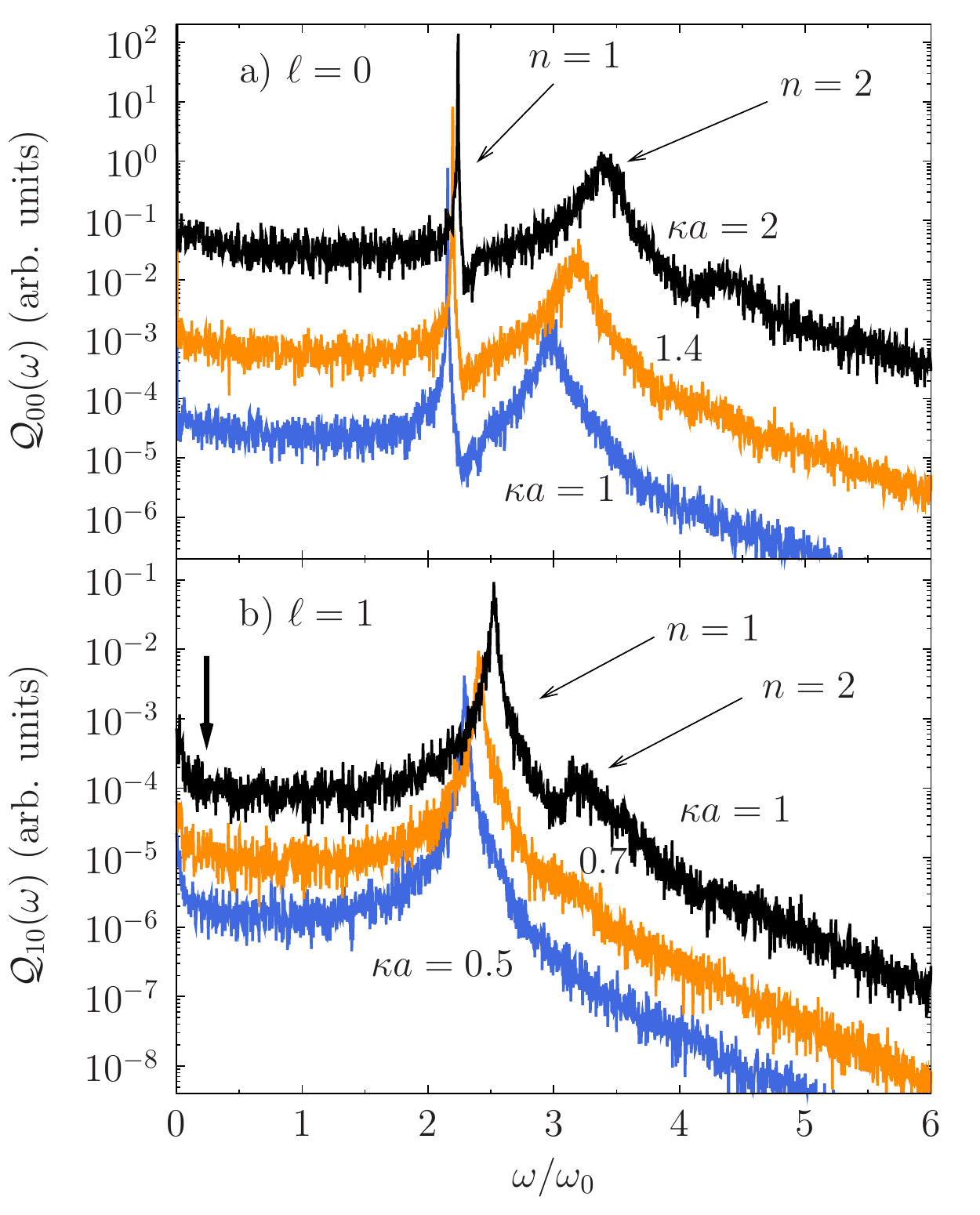}
\caption{Same as Fig.~\ref{fig:kappavar150} for $\Gamma=30$. The thick arrow indicates the position of the low frequency mode for $\Gamma=150$.}\label{fig:kappavar30}
\end{figure}

Comparing the mode frequencies with those for $\Gamma=150$ we find that, on the one hand, they are only weakly affected by the change of $\Gamma$. We still observe good agreement with cold fluid theory as in the previous case. At lower coupling the peak width increases which is indicative of higher damping. On the other hand, there are two important qualitative changes of the spectra. First, the low frequency modes for $\ell=1,2,3$ disappear completely. Second, the modes that exist only for $\xi_{n\ell}>\xi_{n\ell}^\text{crit}$ do not reappear at lower screening. For the monopole and dipole modes this can be seen more clearly by comparing Figs.~\ref{fig:kappavar150} and \ref{fig:kappavar30} which show the same spectra at different coupling strengths. Clearly, the low frequency peak in the dipole spectrum is absent at $\Gamma=30$ (cf. thick arrow in Figs.~\ref{fig:kappavar150}b, \ref{fig:kappavar30}b). Further, the higher order modes with $n>1$ are strongly damped.

\begin{figure}
\includegraphics[width=0.48\textwidth]{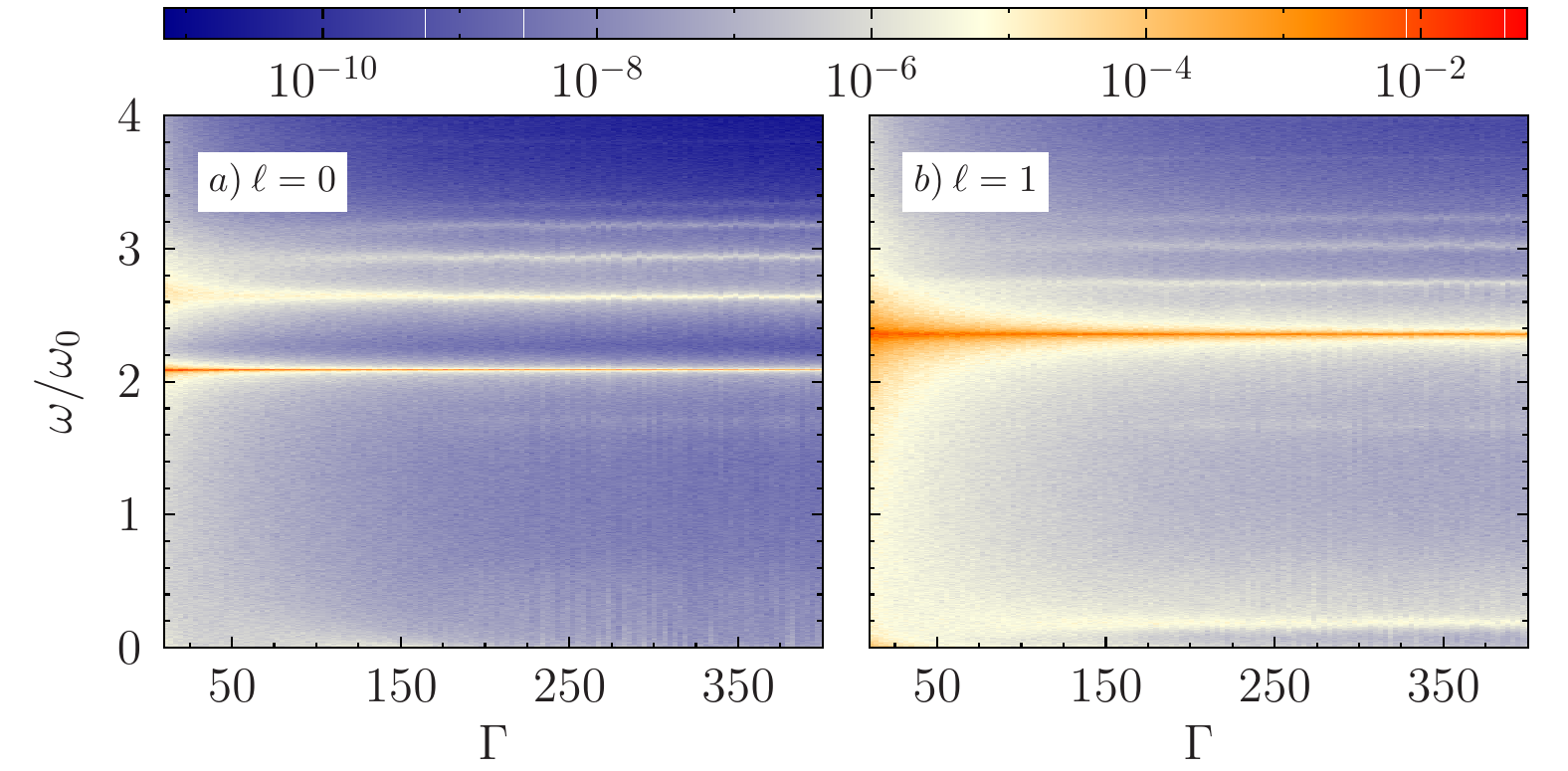}\\
\includegraphics[width=0.48\textwidth]{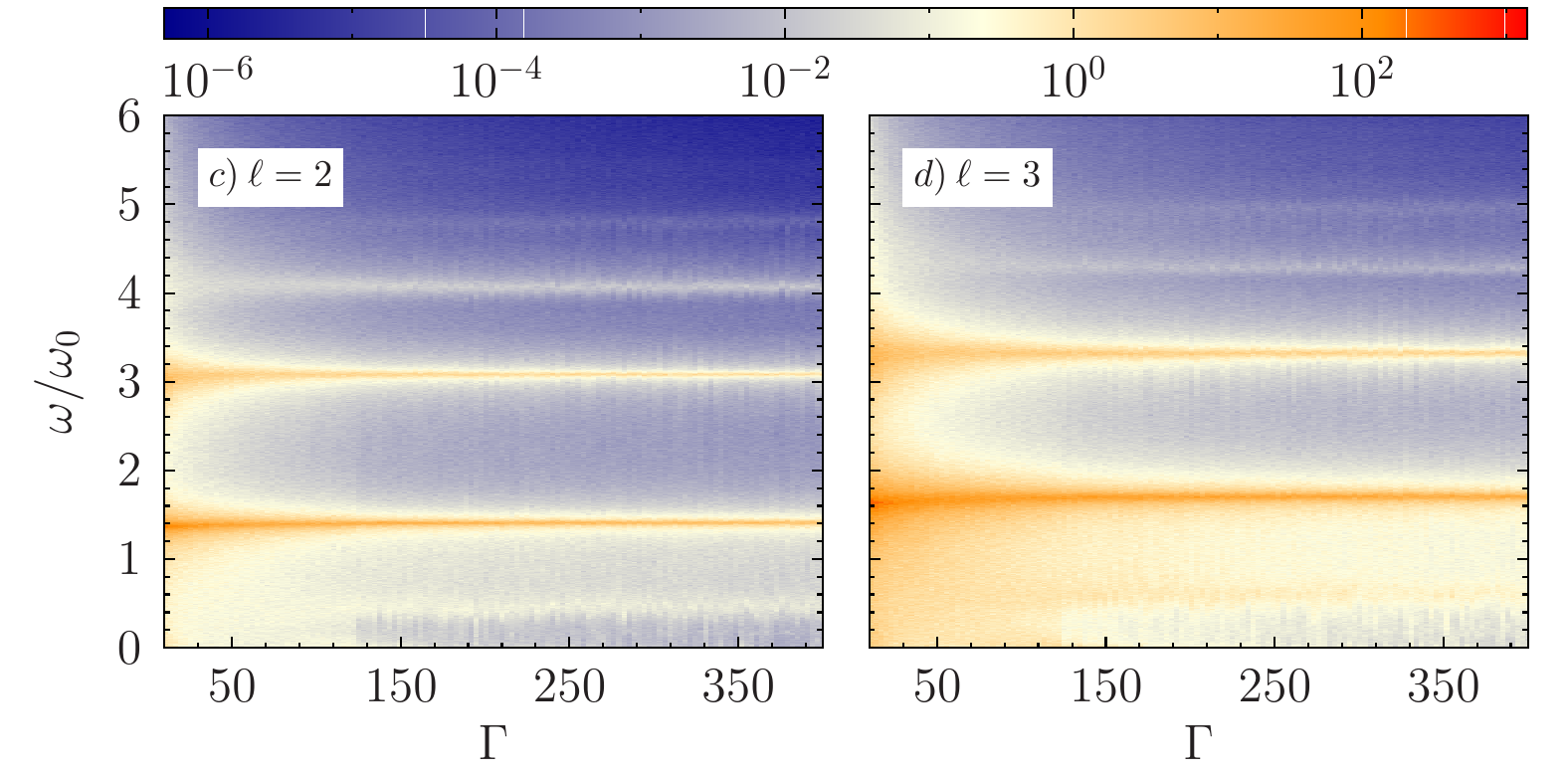}
\caption{Influence of the coupling strength on the $m=0$ multipole spectra $\mathcal Q_{\ell 0}$ (arb. units) for $N=1000$ and $10\le \Gamma\le 400$. The screening parameters are $\kappa a=0.6$ (a,b) and $\kappa a=2$ (c,d).}\label{fig:N1000Gamma}
\end{figure}

\begin{figure}
\includegraphics[width=0.4\textwidth]{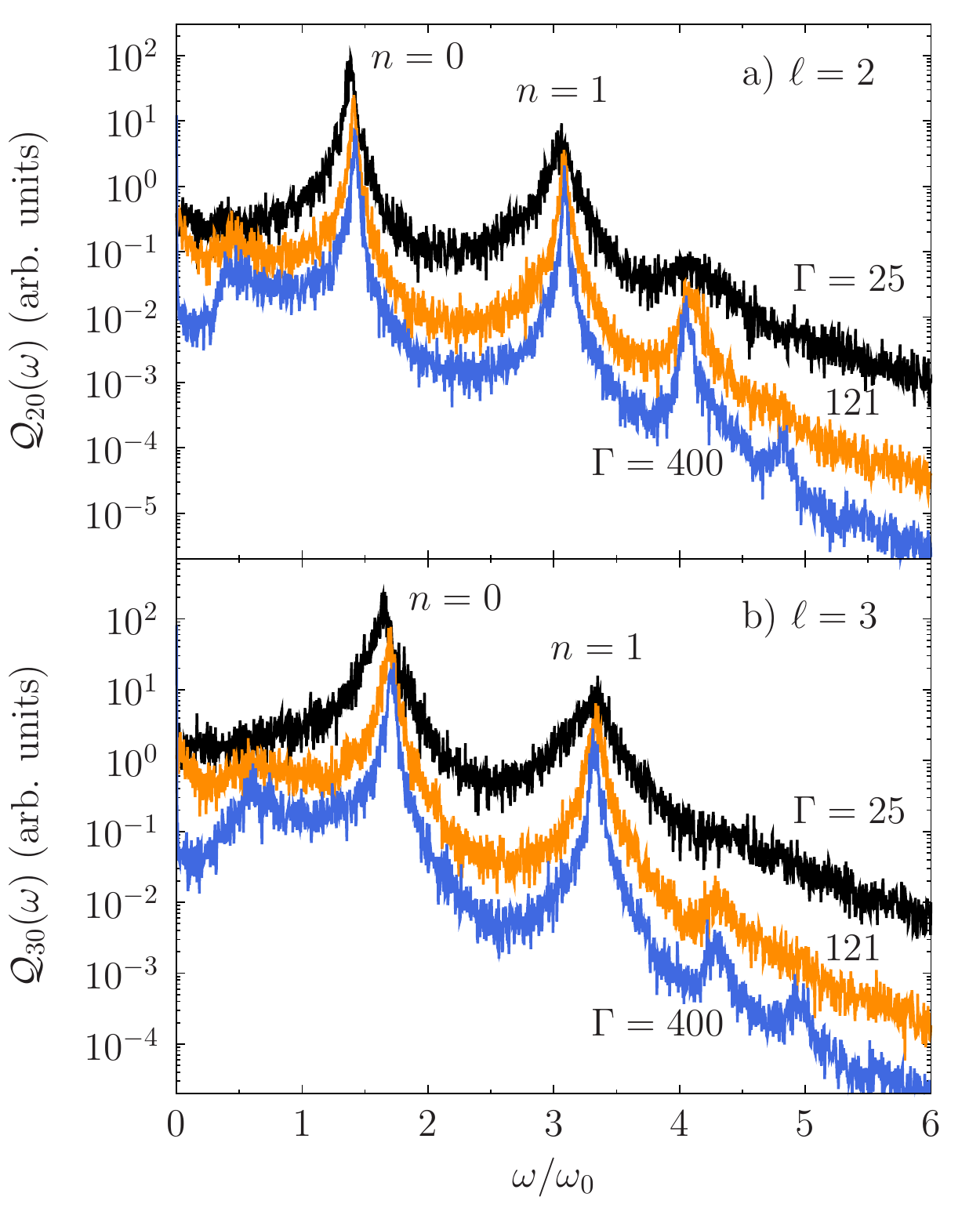}
\caption{Quadrupole and octupole spectra for $N=1000$ and $\kappa a=2$ for three values of the coupling parameter $\Gamma$ as indicated in the figure.}\label{fig:gammavar}
\end{figure}

Figs.~\ref{fig:N1000Gamma} and \ref{fig:gammavar} show the coupling dependence in more detail. Fig.~\ref{fig:N1000Gamma} confirms that the frequencies are almost independent of the coupling parameter, at least in the strong coupling regime. While for $\Gamma>100$ the peaks are well defined, for $\Gamma<100$ the peaks become broader and less pronounced. The vanishing of several modes is clearly visible in Fig.~\ref{fig:gammavar}. At very strong coupling, $\Gamma=400$, the spectrum shows five distinct peaks. Decreasing $\Gamma$ the peaks become broader and the higher order modes vanish. The rather broad maximum at low frequencies is still visible at $\Gamma=121$ but is absent in the $\Gamma=25$ spectrum. The peak positions are only weakly affected by the variation of $\Gamma$. While the frequency of the $n=0$ mode slightly decreases as $\Gamma$ decreases, the mode frequency for $n=1$ increases for $\ell=3$ and decreases for $\ell=2$. We did not investigate this dependence for all modes in further detail because the changes are small and our main interest is in the strongly coupled region with $\Gamma>100$. More details on the $\Gamma$ dependence of the breathing mode can be found in Sec.~\ref{sec:crystalmodes}.

\subsubsection{Influence of damping}
\begin{figure}
\includegraphics[width=0.45\textwidth]{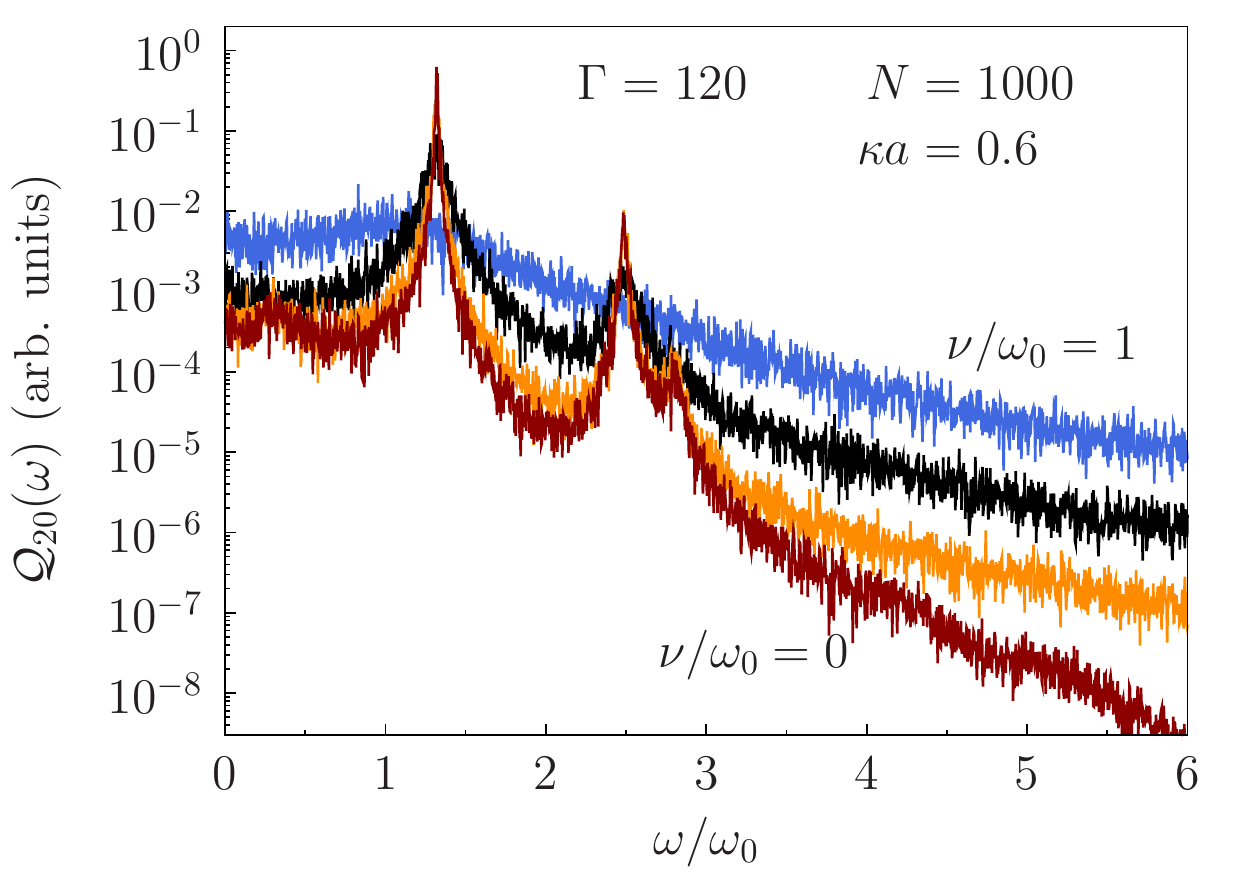}
\caption{Influence of a finite damping coefficient on the quadrupole spectrum for the parameters indicated in the figure. Results are shown for $\nu/\omega_0=0, 0.01,0.1,1$.}\label{fig:nuvar}
\end{figure}

So far our results have been obtained for zero damping. In order to make predictions for experiments a finite damping term must be taken into account. A typical example is shown in Fig.~\ref{fig:nuvar}. For $\nu=0$ the spectrum has two main peaks at $\Omega= 1.32, 2.48$ and two smaller peaks at $\Omega\approx 0.3, 2.8$. With the increase of $\nu$ the peak width increases and their height decreases. At high frequencies we see an increase of the noise level. 

For $\nu/\omega_0=0.001$ all four peaks are still present in the spectrum. At $\nu/\omega_0=0.1$ the two smaller peaks cannot be identified any more and at $\nu/\omega_0=1$ the spectrum is almost flat and only the main peak can be observed. This shows that a low damping rate is essential in order to observe the higher modes with $n=2,3,\dots$ in a Yukawa system. In dusty plasma experiments the friction coefficient can be within a wide range~\cite{ivanov3d,block2008,kroll2010}, depending on the gas pressure. We thus expect that it should be possible to observe the modes in experiments.

\section{Exact crystal normal modes and comparison with fluid theory and MD simulation}\label{sec:crystalmodes}
The collective modes of fluid theory arise from a continuum model of the plasma which completely neglects the underlying discrete particle structure. On the other hand, $N$ strongly coupled correlated particles in a trap possess $3N$ normal modes (related to phonons) which can be directly computed applying harmonic lattice theory. It is, therefore, very interesting to compare the spectra following from fluid theory to the normal mode spectrum.

\subsection{Frequency sum rule}
Before comparing the two approaches (fluid theory and exact normal modes) we discuss a sum rule for the squared eigenfrequencies which is known to be particularly simple in Coulomb systems. The exact normal modes of the discrete $N$-particle system are calculated from the Hessian matrix $\mathcal{U}^s$ of a stationary state $s$ (ground or metastable state). The total potential energy is
\begin{equation}\label{eqn:energys}
 U^s=\sum_{i=1}^{N} V({\vec r}_{i,0}^{s}) + \frac{1}{2}\sum ^N_{i\ne j}  \phi(|\vec{r}_{ij,0}^{s}|),
\end{equation}
where the  $\{\vec r_{i,0}^{s}\}$ are the particle positions. The eigenvalues of $\mathcal{U}^s$ are the (squared) eigenfrequencies of the system, e.g.~\cite{henning08}.

A frequency sum rule for particles with Coulomb interaction in a harmonic trap $V(r)=m\omega_0^2 r^2/2$ is given by $\sum_{i=1}^{3N}(\omega_i^s)^2=3 N\omega_0^2$~\cite{dubin_md96}. Particularly, the sum is independent of the stationary state used to evaluate the Hessian. We will show below that this is not a general rule and does not hold for the Yukawa system considered here.

Since, in general, $\sum_{i=1}^{3N}m(\omega_i^s)^2=\text{Tr}(\mathcal U^s)$ we need to evaluate the diagonal elements of the Hessian,
\begin{equation}\label{eqn:sumrule}
\text{Tr}(\mathcal U)=\sum_{i=1}^{N} \Delta_i  \left[ \sum_{k=1}^NV(\vec r_k) + \frac{1}{2}\sum_{j\ne k}^N \phi(|\vec r_{jk}|) \right],
\end{equation}
at $\vec r_i=\vec r_{i,0}^s$. For the harmonic confinement considered here we have $\Delta V(r)=3m\omega_0^2$. The interaction term is simply the radial part of the Laplacian applied to $\phi(r)$. As the Yukawa potential satisfies $(\Delta-\kappa^2) e^{-\kappa r}/r=-4\pi \delta(\vec r)$, the result is
\begin{equation}\label{eqn:sumruleYukawa}
 \sum_{j=1}^{3N}\left(\frac{\omega_j^s}{\omega_0}\right)^2=3N + 2\,(\kappa a)^2 \,\frac{U^s_{\text{int}}}{q^2/a},
\end{equation}
 where $U^s_\text{int}$ denotes the total interaction energy of the particles in the state $s$ [second term in Eq.~(\ref{eqn:energys})]. The delta function does not contribute since $|\vec r_{ij,0}^s|\ne 0$.

For $\kappa=0$ Eq.~(\ref{eqn:sumrule}) reduces to the known Coulomb limit. Due to the additional interaction term the frequency sum for Yukawa interaction explicitly depends on the particle configuration, e.g. the shell occupation numbers or the arrangement of the particles on a given shell. It is not a universal quantity like in the Coulomb case. Eq.~(\ref{eqn:sumruleYukawa}) connects the eigenfrequencies of the state $s$ with its interaction energy, i.e. it establishes a connection between dynamic and static properties. We computed both sides of Eq.~(\ref{eqn:sumruleYukawa}) independently and found that the sum rule is indeed fulfilled.

\subsection{Comparison of MD simulation with crystal eigenmodes}
In the crystalline phase the particle motion is mainly determined by the crystal eigenmodes and, consequently, the same applies to the oscillations of the multipole moments. In the following we extract the dominant contribution to the multipole moments by identifying the eigenmode that is closest to the respective fluid mode. This is achieved by computing the particle displacements according to cold fluid theory and projecting all $3N$ eigenmodes on the so-defined reference mode, see also~\cite{dubin_md96}. This is done according to
\begin{align*}
 c_j = \left(\frac{\vec w(\vec r_1,\dots,\vec r_N)}{|\vec w(\vec r_1,\dots,\vec r_N)|}\cdot\vec v_j \right)^2,
\end{align*}
where $\vec v_j$ is the $j$th (normalized) eigenvector of the stationary state of the crystal with particle positions $\{\vec r_{i,0}^s\}$ and $\vec w$ is the eigenvector calculated from cold fluid theory. The particle displacements for the monopole modes in this case are $\vec w_i= \left.f'(r,\omega) \hat e_r\right|_{\vec r=\vec r_{i,0}^s}$. The full displacement vector is then taken as $\vec w= (\vec w_1,\dots,\vec w_N)$. The mode with the greatest projection is chosen for comparison with the simulation data. The stationary states used for the calculation of the normal modes were found by slowly cooling a random initial state in a molecular dynamics simulation until the equilibrium positions were reached, e.g.~\cite{Patrick2005,kaehlert2008}. This procedure was repeated 200 times for each parameter set ($N,\kappa a$). The state with the lowest energy was then chosen to evaluate the Hessian.

\begin{figure}
\includegraphics[width=0.48\textwidth]{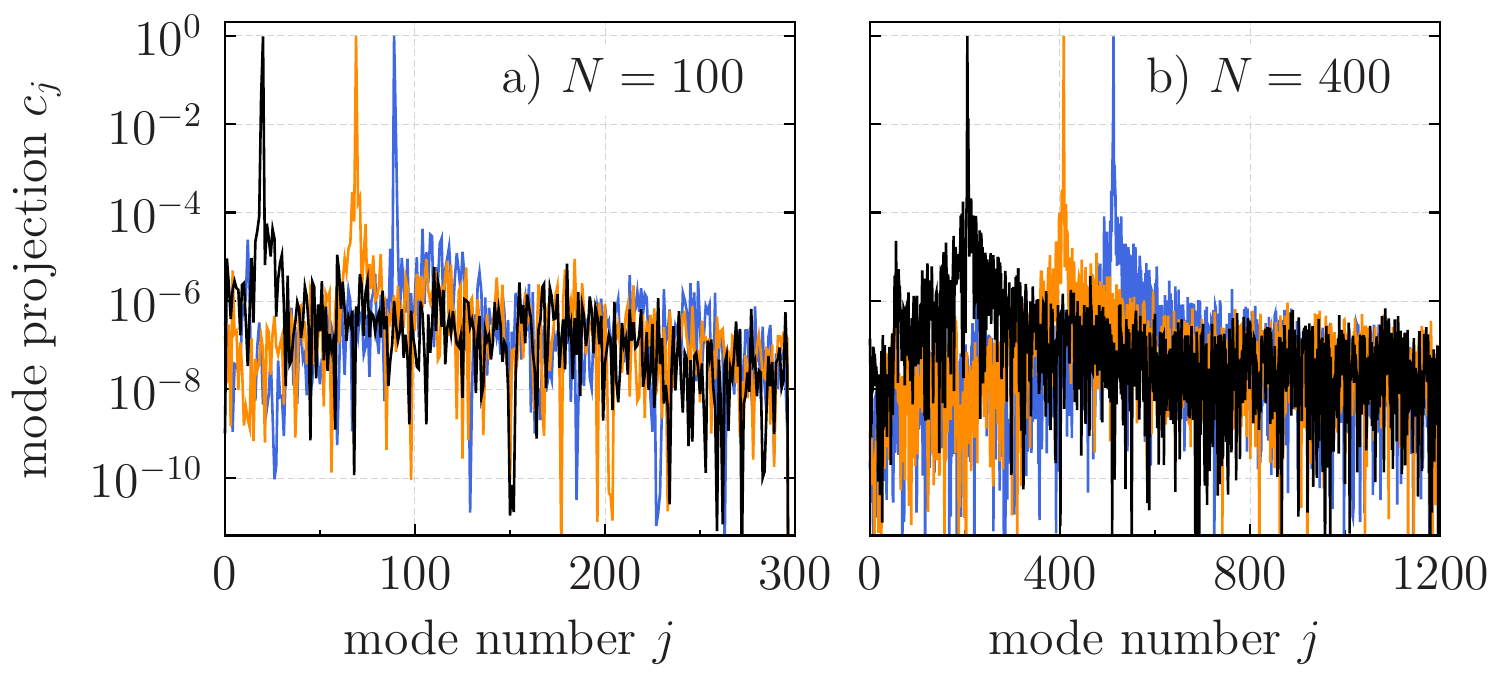}
\caption{Projection of the $n=1$ monopole mode ($\ell=m=0$) of fluid theory on the $3N$ crystal eigenmodes. Results are shown for $\kappa a =0.4, 1.2, 2$ (from left to right peak). The modes are ordered by their frequency (from high to low).}\label{fig:n1L0projection}
\end{figure}

The projections on the monopole mode with $n=1$ are shown in Fig.~\ref{fig:n1L0projection}. One can clearly see that only very few modes have a non-negligible projection on this fluid mode. There is a single mode with a high projection - this corresponds to a 'breathing mode'. While for low $\kappa a$ the breathing mode is among the normal modes with the highest frequencies, the peak shifts to intermediate frequencies as screening increases. This is very similar to the behavior in cold fluid theory where additional modes appear at larger screening which have frequencies higher than that of the breathing mode, cf. Fig.~\ref{fig:N1000G150}. In contrast, for Coulomb interaction the breathing mode is always the one with the highest frequency~\cite{dubin_md96}.

\begin{figure}
\includegraphics[width=0.48\textwidth]{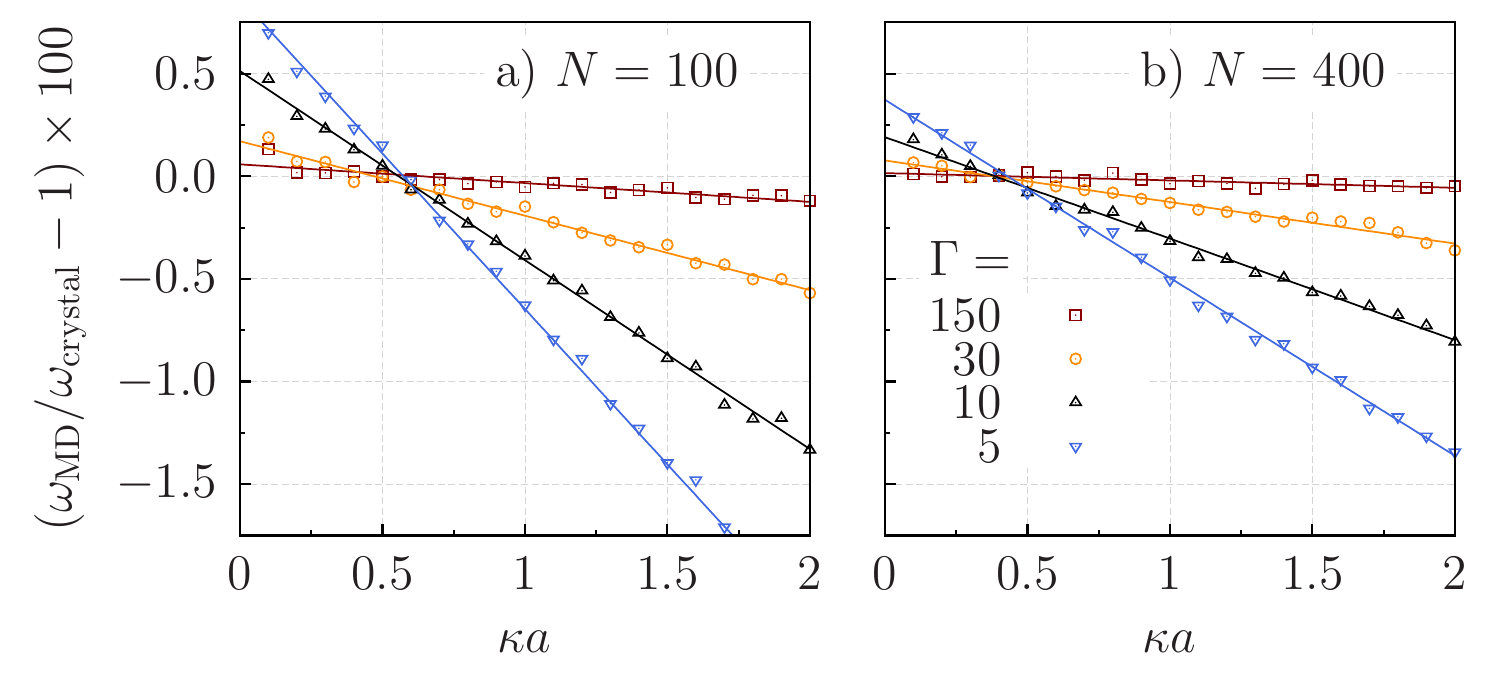}
\caption{Deviation of the $n=1$ monopole mode from the frequency of the exact eigenmode with the greatest projection on the fluid mode for a) $N=100$ and b) $N=400$ at $\Gamma=5,10,30,150$. The solid lines show a linear fit.}\label{fig:n1L0MD}
\end{figure}

The very narrow peak of the breathing mode in the MD simulations allows for a detailed investigation of its frequency. In Fig.~\ref{fig:n1L0MD} we compare the frequency obtained in the MD simulation with the frequency of the projected crystal eigenmode [peak in Fig.~\ref{fig:n1L0projection}].  While for low $\kappa a$ the frequency in the simulation exceeds the eigenfrequency of the crystal, the trend is reversed for larger $\kappa a$. As the coupling parameter is increased the deviations between the two frequencies decrease and vanish almost identically in the very strong coupling limit, $\Gamma=150$. 

The increase of the deviations with temperature (reduction of $\Gamma$) is readily understood by the limitations of harmonic lattice theory: when $T$ increases, the amplitude of particle fluctuations around their stationary positions grows and cannot be described by the second order contribution in the expansion of the potential energy. Anharmonic corrections and mode coupling effects (which are contained in the MD simulation) lead to deviations from the normal modes of the harmonic approximation. The interesting observation following from Fig.~\ref{fig:n1L0MD} is that the sign of these deviations changes with $\kappa a$. We note that we also compared the breathing frequencies of different stationary states for a few cases and found that the typical difference is of the order $\Delta \Omega\sim 10^{-4}$ and therefore about an order of magnitude smaller than those caused by the finite temperature in the MD simulation.

The observed temperature dependence is peculiar: At low screening the MD mode frequency decreases with an increase of $\Gamma$ while it increases at larger $\kappa a$. This behavior is in qualitative agreement with the results of~\cite{olivetti2009} who found that the breathing frequency increases (decreases) with the coupling strength for short- (long-) ranged forces of the power law type. Even though their analysis does not directly apply to our results with Yukawa interaction the general trend is the same. In our system the interaction range is essentially determined by the parameter $\xi=\kappa R$. In a very small interval around $\kappa a\approx 0.6$ ($N=100$) and $\kappa a\approx 0.4$ ($N=400$) the breathing frequency is almost independent of $\Gamma$.  Here $\xi\approx 2.3$ and $\xi\approx 2.4$, respectively, which could be an indication that the temperature dependence of the breathing frequency in the strongly coupled regime is essentially determined by the value of $\xi$.

\begin{figure}
\includegraphics[width=0.48\textwidth]{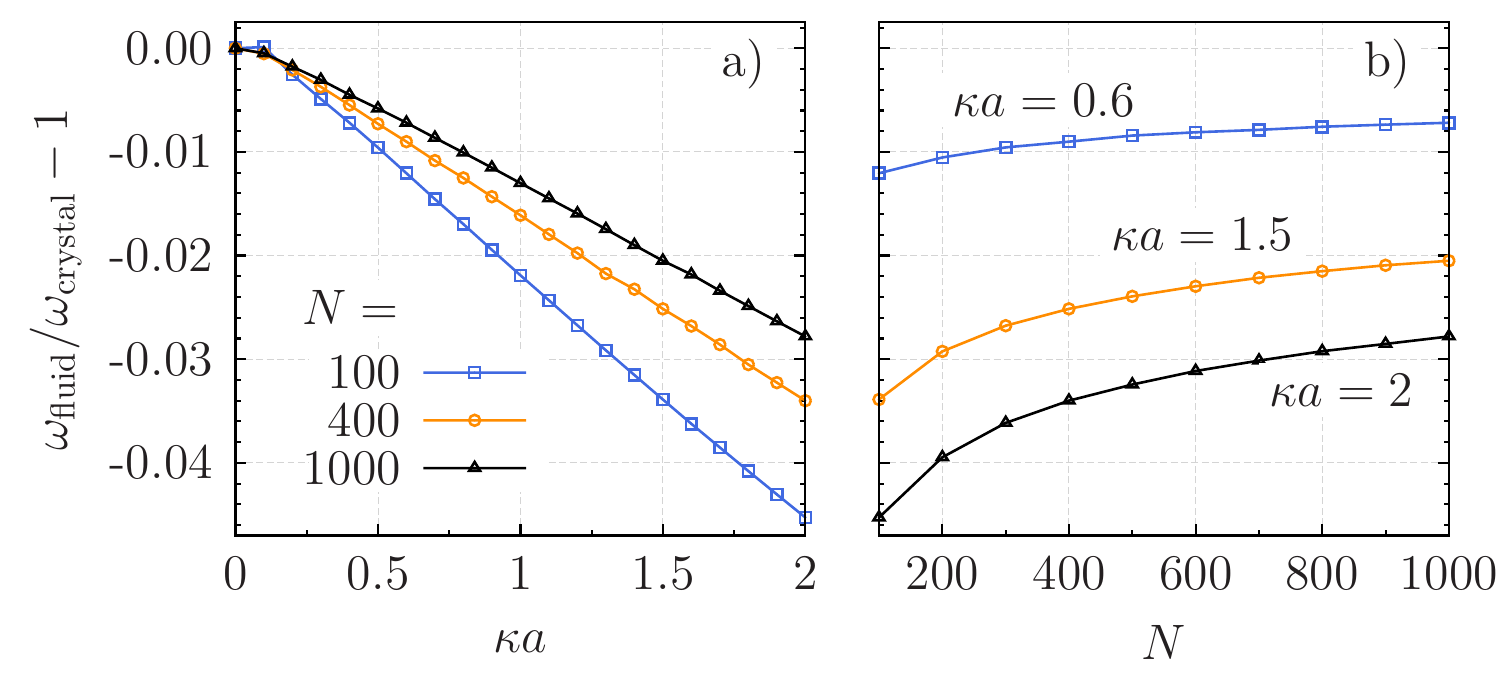}
\caption{Frequency deviation of the exact eigenmode with the greatest projection on the fluid mode from the results of cold fluid theory as a function of a) screening and b) number of particles.}\label{fig:deviation}
\end{figure}

\subsection{Comparison of fluid theory with crystal eigenmodes}
The previous analysis showed that the frequency in the MD simulation approaches the frequency of the crystal eigenmode in the very strong coupling regime. In order to test the validity of the fluid theory, which was derived for the infinite coupling case ($\Gamma\to\infty$), we, therefore, compare its frequencies with those of the exact crystal eigenmodes. This has two advantages over the MD simulation: 1) the mode frequency can be determined with a high accuracy and 2) the infinite coupling condition is already 'built-in'.

The results are shown in Fig.~\ref{fig:deviation}. In the Coulomb limit the breathing mode is among the bulk modes with the (exact) frequency $\Omega=\sqrt{3}$ and uniform particle displacements~\cite{henning08}. Even though the particle displacements in the $n=1$ monopole mode for $\xi\to 0$ in cold fluid theory do not become uniform~\cite{kaehlert2010}, the mode with the greatest projection is nevertheless the breathing mode, and so the frequency deviations in the Coulomb limit vanish, at least for the particle numbers considered here. For finite screening no universal uniform breathing mode exists~\cite{henning08}. Here we compared the projection according to the theory with uniform displacements, i.e. $\vec w_i= \left.r \hat e_r\right|_{\vec r=\vec r_{i,0}^s}$, and found that both methods select the same modes (with an exception for $N=100$, $\kappa a=0.1$). In almost all cases the projection in the latter case is slightly less than in the former, i.e. the mode form is more accurately described by non-uniform particle displacements.

The deviations of the mode frequencies are of the order of a few percent and about an order of magnitude larger than those caused by finite temperature effects, cf. Figs.~\ref{fig:n1L0MD} and~\ref{fig:deviation}. They increase linearly with $\kappa a$ and decrease with $N$. It was shown in~\cite{sheridan2006} that the breathing frequency increases if correlations are included in the theory. This explains why the fluid treatment underestimates the mode frequencies. Our results suggest that in the limit $N\to\infty$ a crystal eigenmode could exist that directly corresponds to the $n=1$ monopole fluid mode. Its frequency would be given by $\Omega=\sqrt{5}$ [cf. Eq.~(\ref{eqn:freqlimit}) for $(n,\ell)=(1,0)$], in contrast to the Coulomb limit where $\Omega=\sqrt{3}$.

\section{Conclusion}\label{sec:conclusion}
In this paper we have analyzed the normal modes of a spherically confined Yukawa plasma over a broad range of coupling parameters $\Gamma$ and screening parameters $\kappa a$. This question was studied using three different approaches: MD simulations, harmonic lattice theory (crystal normal modes) and a recently developed fluid theory. Of central interest was to test the predictions of the latter: the existence of an additional (compared to Coulomb systems) class of collective modes enumerated by the mode number $n$.

Our first conclusion is that 1) these modes are also present in the exact results. 2) The lowest order mode ($n=0$) of fluid theory practically coincides with the MD result for all studied cases. 3) With increasing mode index $n\ge 1$ the frequencies predicted by fluid theory are systematically too low whereas the minimal value $\kappa a$ where the modes exist is too large. 4) The results of fluid theory become more accurate when the particle number increases, in agreement with the continuum character of the approach. 5) Comparison of the fluid monopole mode with $n=1$ to the breathing mode of crystal harmonic lattice theory showed that the main source of error are missing correlations: The frequency of fluid theory is systematically too small and deviations increase with $\kappa a$. 6) Stimulated by the predictions of fluid theory we have performed detailed MD simulations which have uncovered the complete collective excitation spectrum. In addition to the fluid theory we observed two peculiarities which are present at strong coupling: i) a low frequency mode below the $n=0$ modes of fluid theory and ii) reentrant character of fluid modes (which exist beyond a critical value of $\kappa $) at low values of $\kappa a$. The physical origin of these effects is still unclear and will be explored elsewhere.

Since the effects of a finite temperature on the mode frequencies is negligible when compared with those caused by correlations, an improved fluid theory should mainly be concerned with the inclusion of correlations. Correlation corrections for the breathing mode in Yukawa systems were accounted for by introducing a finite particle separation~\cite{sheridan2006}. Recently, Olivetti et al.~\cite{olivetti2009} derived an expression for the breathing frequency of particles with power law interaction which accounts for the effects of finite temperature and correlations. For Coulomb systems the effects of finite temperature and correlations were investigated in~\cite{dubin_corr96}. An improved theory for the modes investigated here is subject of ongoing work.

Finally, since MD simulations were recently found to accurately reproduce the properties of Yukawa balls~\cite{bonitz2006,block2008,kaeding,kaehlert2008,bonitz2010rev} we expect that the reported excitation spectrum should be directly observable in dusty plasma experiments provided the damping rate is sufficiently low.

\acknowledgements
This work is supported by the DFG via SFB-TR24 and the North-German Supercomputing Alliance (HLRN).

\end{document}